\documentstyle[12pt]{article}

\advance\textheight by \topskip
\begin{document}
\tolerance=100000
\def\Dir{\kern -6.4pt\Big{/}}
\def\DDir{\kern -7.6pt\Big{/}}
\def\DGir{\kern -6.0pt\Big{/}}
\def\CCl{\kern -6.6pt^{(-)}}
\def\Ord{\buildrel{\scriptscriptstyle <}\over{\scriptscriptstyle\sim}}
\def\simlt{\rlap{\lower 3.5 pt \hbox{$\mathchar \sim$}} \raise 1pt \hbox {$<$}}
\def\OOrd{\buildrel{\scriptscriptstyle >}\over{\scriptscriptstyle\sim}}
\def\simgt{\rlap{\lower 3.5 pt \hbox{$\mathchar \sim$}} \raise 1pt \hbox {$>$}}
\def\sqr#1#2{{\vcenter{\hrule height.#2pt
      \hbox{\vrule width.#2pt height#1pt \kern#1pt
        \vrule width.#2pt}
      \hrule height.#2pt}}}
\def\square{\mathchoice\sqr68\sqr68\sqr{4.2}6\sqr{3.0}6}
\def\leaderfill{\leaders\hbox to 1em{\hss.\hss}\hfill}
\font\fivesy=cmsy5 
\font\tensy=cmsy10 
\def\pscal{\hbox{\hskip2pt\fivesy\raise .5ex\hbox{\char'017}\hskip2pt }}
\def\pvet{\hbox{\hskip2pt\tensy\char'002\hskip -7.3pt\char'002\hskip2pt}}
\def\pmb#1{\setbox 0=\hbox{$#1$}%
\kern-0.25em\copy0\kern-\wd0
\kern0.5em\copy0\kern-\wd0
\kern-0.25em\raise.0433em\box0}
\def\epem{\ifmmode{e^+ e^-} \else{$e^+ e^-$} }
\def\Im{\mathop{{\cal I}\mskip-4.5mu \lower.1ex \hbox{\it m}}}
\def\Re{\mathop{{\cal R}\mskip-4mu \lower.1ex \hbox{\it e}}}
\def\pl #1 #2 #3 {{\it Phys.~Lett.} {\bf#1} (#2) #3}
\def\np #1 #2 #3 {{\it Nucl.~Phys.} {\bf#1} (#2) #3}
\def\zp #1 #2 #3 {{\it Z.~Phys.} {\bf#1} (#2) #3}
\def\pr #1 #2 #3 {{\it Phys.~Rev.} {\bf#1} (#2) #3}
\def\prep #1 #2 #3 {{\it Phys.~Rep.} {\bf#1} (#2) #3}
\def\prl #1 #2 #3 {{\it Phys.~Rev.~Lett.} {\bf#1} (#2) #3}
\def\mpl #1 #2 #3 {{\it Mod.~Phys.~Lett.} {\bf#1} (#2) #3}
\def\rmp #1 #2 #3 {{\it Rev. Mod. Phys.} {\bf#1} (#2) #3}
\def\xx #1 #2 #3 {{\bf#1}, (#2) #3}
\def\preprint{{\it preprint}}
\def\sm{${\cal {SM}}$}
\def\mssm{${\cal {MSSM}}$}
\input wrl$grp04:[moretti.feynman]feynman
\thispagestyle{empty}
\setcounter{page}{0}

\begin{flushright}
{\large DFTT 79/93}\\
{\rm December 1993\hspace*{.5 truecm}}\\
\end{flushright}

\vspace*{\fill}

\begin{center}
{\Large \bf Intermediate mass Higgs bosons at TeV $e\gamma$ colliders
in the ${\cal {MSSM}}$\footnote{Work supported in part by Ministero
dell' Universit\`a e della Ricerca Scientifica.}}\\[2.cm]
{\large Stefano Moretti}\\[.3 cm]
{\it Dipartimento di Fisica Teorica, Universit\`a di Torino,}\\
{\it and INFN, Sezione di Torino,}\\
{\it V. Pietro Giuria 1, 10125 Torino, Italy\footnote{Address after
January 1994: {\it Dept. of Physics, University of Durham,
South Road, Durham DH1 3LE, England, U.K.}}}.\\[2cm]
\end{center}

\vspace*{\fill}

\begin{abstract}
{\normalsize
\noindent
We study the cross sections for
the production of intermediate mass Higgs bosons
via the reactions
$e^-\gamma\rightarrow \nu_eW^-\Phi^0$,
$e^-\gamma\rightarrow \nu_eH^-\Phi^0$,
$e^-\gamma\rightarrow e^-Z^0\Phi^0$ and
$e^-\gamma\rightarrow e^-H^+H^-$
in the Minimal Supersymmetric Standard Model
(with $\Phi^0=H^0,h^0$ and, where possible, $A^0$) at TeV energies,
taking into account $b$--tagging capabilities.
We find that the rates
for $e^-\gamma\rightarrow \nu_eW^-\Phi^0$
are large enough to permit the discovery of at least one
between $h^0$
and $H^0$ over the whole intermediate mass range of $M_{A^0}$,
at all $\tan\beta$, using the $(jj)(b\bar b)$ final state.
The ${\cal {CP}}$-odd neutral $A^0$ and the charged
$H^\pm$'s can be detected via the processes
$e^-\gamma\rightarrow \nu_eH^-A^0$,
with $A^0\rightarrow b\bar b$, and
$e^-\gamma\rightarrow e^-H^+H^-$, resorting
to the leptonic decay
$H^\pm\rightarrow \nu_\tau\tau^+(\bar\nu_\tau\tau^-)$,
but only for $M_{A^0}\approx60$ GeV.
Finally, we present explicit formulae for the helicity amplitudes
of these processes.}
\end{abstract}

\vspace*{\fill}

\newpage
\subsection*{Introduction}

The Higgs mechanism which is assumed to break the electroweak
$SU(2)_L\times U(1)_Y$ gauge symmetry spontaneously
has not been yet tested by the experiments.\par
The simplest version is the Standard Model (\sm), where
the $SU(2)_L$ symmetry is broken by a doublet of fundamental scalar
fields with a non--zero vacuum expectation value.
In the Minimal Supersymmetric Standard Model (\mssm) this
happens by means of two doublets.
These models predict the existence of scalar particles,
one in the case of the \sm, the ${\cal {CP}}$--even neutral $\phi$,
and five in the \mssm, the ${\cal {CP}}$--even neutral ones $H^0$ and $h^0$,
the ${\cal {CP}}$--odd neutral one $A^0$ and the charged ones
$H^\pm$'s\footnote{The
three neutral Higgs states of the \mssm\ will be collectively indicated by the
symbol $\Phi^0$.}.
Lower limits on their masses can be extracted at present
colliders. From LEP I ($\sqrt s_{ee}=M_{Z^0}$) experiments,
by the results of searches for $Z^{0*}\phi$ events, one derives the bound
\begin{equation}
M_\phi\OOrd 62.5~{\mathrm {GeV}},
\end{equation}
for the ${\cal {SM}}$ Higgs \cite{limSM}.
Using the reactions $e^+e^-\rightarrow Z^{0*}h^0$ and
$e^+e^-\rightarrow h^0A^0$, the lower limits on ${\cal {MSSM}}$
neutral Higgses\footnote{For the typical choice of parameters
$m_t=150$ GeV and $m_{\tilde t}=1$ TeV, see later on.}
are presently \cite{limMSSM}
\begin{equation}
M_{h^0}\OOrd 44.5~{\mathrm {GeV}}\quad\quad{\mathrm {and}}\quad\quad
M_{A^0}\OOrd
45~{\mathrm {GeV}}.
\end{equation}
Wide studies have been carried out on the detectability of
a Higgs particle by the next generation of high energy
machines, both at a $pp$ hadron collider \cite{guide,LHC,SSC} and at an
$e^+e^-$ Next Linear Collider (NLC) \cite{guide,LepII,NLC,ee500,JLC}.\par
The regions $M_{\phi,A^0} < 80-90$ GeV will be studied at LEP II
($\sqrt s_{ee}=160-200$ GeV) by the Higgs decay channel $b\bar b$
\cite{LepII}, for the \sm\ via the Bjorken bremsstrahlung reaction  $e^+e^-
\rightarrow Z^{0*}\rightarrow Z^0\phi$ \cite{bremSM}, and for the \mssm\ via
one or both the processes
$e^+e^-\rightarrow Z^{0*}\rightarrow Z^{0}h^0$ (bremsstrahlung) and
$e^+e^-\rightarrow Z^{0*} \rightarrow h^0A^0$ (neutral pair production)
\cite{brenppMSSM}.\par
Higgses with larger masses will be searched for
at $pp$ colliders like LHC(SSC), with $\sqrt s_{pp}=16(40)$ TeV and
${\cal L}\approx 100(10)$ fb$^{-1}$,
or at $e^+e^-$ NLCs, with $\sqrt s_{ee}=300\div2000$ GeV and
${\cal L}\approx 10\div20$ fb$^{-1}$.\par
At LHC/SSC, because of the huge QCD background,
the mass range 80 GeV $\Ord M_{\phi,\Phi^0}
\Ord 130$ GeV results the most difficult to study since
in this case a neutral Higgs boson mainly
decays to $b\bar b$ pairs, both in the \sm\ and,
for a large choice of parameters, also in the \mssm.
However, recent studies have shown that the discovery of a \sm\ Higgs $\phi$
is possible by the $\gamma\gamma$ decay mode \cite{gamgam},
via the associated production with a $W^\pm$ boson \cite{gny,wh}
or a $t\bar t$ pair \cite{rwnz,tth}. For $M_\phi \OOrd 130$ GeV,
the ``gold-plated'' four--lepton mode $\phi\rightarrow Z^0Z^0\rightarrow
\ell\bar\ell\ell\bar\ell$
guarantees, in general,
the detection of the \sm\ Higgs \cite{LHC,SSC}.\par
At $\sqrt s_{ee}=300-500$ GeV NLCs it results viable the discovery of the \sm\
Higgs $\phi$
by a large variety of channels over the whole intermediate mass range
\cite{BCDKZ},
both via the bremsstrahlung mechanism and via the fusion processes
$e^+e^-\rightarrow \bar\nu_e\nu_eW^{\pm*}
W^{\mp*}(e^+e^-Z^{0*}Z^{0*})\rightarrow\bar\nu_e\nu_e
(e^+e^-)\phi$ \cite{fusionSM}. A heavy Higgs $\phi$, other than via the $4\ell$
mode,
can be detected at $\sqrt s_{ee}=500$ GeV also via the four--jet mode
$\phi\rightarrow W^\pm W^\mp,Z^0Z^0\rightarrow jjjj$ \cite{BCKP,4jet}.\par
In the \mssm\ the $\Phi^0\rightarrow\gamma\gamma$ mode at hadron colliders
can be exploited for the discovery of $H^0$
for 80 GeV $\Ord M_{A^0} \Ord$ 100 GeV
and of $h^0$ for $M_{A^0}\OOrd$ 170 GeV, at all $\tan\beta$;
while the ``gold--plated'' decay channel is useful for the $H^0$
with $\tan\beta\Ord 7$ and 100 GeV $\Ord M_{A^0} \Ord$ 300 GeV, but not
for the $h^0$ because of its too light mass\footnote{For the $stop$ mass
$m_{\tilde t}=1$ TeV
and all --ino masses greater than 200 GeV.}.\par
Recently, it has been shown \cite{btagg} that with the $b$--tagging
capabilities \cite{SDC} of LHC/SSC experiments\footnote{For LHC this is true
only
if the higher luminosity and larger number of tracks per event can successfully
be dealt with.} it may be possible to detect a \sm\
Higgs boson in the $t\bar t\phi$ production channel, with one $t$
decaying semileptonically and $\phi\rightarrow b\bar b$, for 80 GeV $\Ord
M_\phi \Ord$ 130 GeV,
provided that $m_t\OOrd 130$ GeV.
This channel results useful, over a substantial portion of the parameter space,
for at least one of the ${\cal {MSSM}}$ Higgses $h^0$ or $H^0$, too,
removing the ``window of inobservability'' for
100 GeV $\Ord M_{A^0} \Ord$ 170 GeV and  $\tan\beta\OOrd 2$,
which remained in previous analyses.
Moreover, it has been found \cite{noiA0} that also the reaction $bg\rightarrow
bZ^0\Phi^0$ is an excellent candidate for the discovery
of $A^0$ and at least one of the other two neutral Higgses
over the whole intermediate range of $M_{A^0}$ for
large values of $\tan\beta$, through the same decay channel $\Phi^0\rightarrow
b\bar b$. With regard to charged Higgses, for lower(higher) values of
$M_{H^\pm}$
the dominant production mechanism is $gg\rightarrow t\bar t\rightarrow
H^+H^-b\bar b$($bg\rightarrow tH^-$). Because of QCD backgrounds
only the lower mass case gives a detectable signal over a non--negligible
region
of $(M_{A^0},\tan\beta)$ \cite{kz}.\par
At NLCs energies, other than via bremsstrahlung and neutral pair production
(this latter for $H^0A^0$ final states, too \cite{brenppMSSM}),
\mssm\ Higgses can be produced also via the fusion processes
$e^+e^-\rightarrow \bar\nu_e\nu_eW^{\pm*}
W^{\mp*}(e^+e^-Z^{0*}Z^{0*})\rightarrow\bar\nu_e\nu_e
(e^+e^-)h^0/H^0$ \cite{fusionMSSM} and the charged pair production
$e^+e^-\rightarrow\gamma^*,Z^{0*}
\rightarrow H^+H^-$ \cite{cppMSSM}. The lightest ${\cal {CP}}$--even
Higgs $h^0$ can be detected over the whole \mssm\ parameter space,
independently of the $top$ and
$squark$ masses. Therefore, if the $h^0$ will not be found, the \mssm\ is ruled
out.
If the $H^0$ and $A^0$ boson masses are less then $\approx 230$ GeV, it exists
a very
large area in the parameter space where all neutral Higgses can be
contemporaneously
detected for $\sqrt s_{ee}=500$ GeV \cite{DHZ}.
A charged Higgs with $M_{H^\pm}<m_b+m_t$
mainly decays to $\nu_\tau\tau^+(\bar\nu_\tau\tau^-)$ and $c\bar s(\bar c s)$
pairs (with the leptonic mode dominating
for $\tan\beta>1$). If kinematically allowed, a heavy $H^\pm$ decays via
the $top$ mode $H^\pm\rightarrow t\bar b(\bar t b)$ (and in some part of the
parameter space also
to $W^\pm h^0$). In both cases the signature is a cascade with a $\tau$ or
a $b$ in the final state: so, an extremely good mass resolution is crucial in
order
to reduce the backgrounds from $top$ and boson pair production. For an
intermediate $H^\pm$, if $\tan\beta>1$,
a possible signature is an apparent breaking of the $\tau$ {\it vs.}
$\mu$/$e$ universality.\par
At higher $e^+e^-$ energies, such as $\sqrt s_{ee}=$1--2 TeV, fusion
mechanisms become dominant over other production processes,
both in the \sm\ and in the \mssm\ \cite{JLC,Bur}.\par
The conversion of linear
$e^+e^-$ NLCs into $\gamma\gamma$ and/or $e\gamma$ colliders, by
photons generated via Compton back--scattering of laser light,
provides new possibilities of detecting and studying Higgs bosons
\cite{laser}.\par
An important channel, by $\gamma\gamma$ collisions at $\sqrt s_{ee}=500$ GeV,
is the production of a heavy \sm\ Higgs  $\phi$  (up to $\approx 350$ GeV)
by a triangular loop of heavy fermions or $W^\pm$, with the detection via the
decay mode $\phi\rightarrow
Z^{0}Z^{0}\rightarrow q\bar q\ell^+\ell^-$ \cite{heavyphphSM}. Moreover,
the process $\gamma\gamma\rightarrow
t\bar t \phi$ results more useful than the corresponding $e^+e^-$ process
in measuring the Yukawa coupling $t\phi$ at $\sqrt s_{ee}=1$--2 TeV
\cite{Yukawa}.\par
For the \mssm, at a NLC with $\sqrt s_{ee}=500$ GeV,
$\gamma\gamma\rightarrow\Phi^0$ reactions are important
in searching for heavy $H^0$ and $A^0$ bosons: they
can be detected up to mass values of $\approx 0.8\sqrt s_{ee}$, for moderate
$\tan\beta$ and if a luminosity of 20 fb$^{-1}$, or more, can be achieved
\cite{gMSSM}.
For the $H^0$, the channels $H^0\rightarrow h^0h^0$, if $M_{H^0}
\Ord 2m_t$, and $H^0\rightarrow t\bar t$, for $M_{H^0}\OOrd 2m_t$, appear more
interesting than the decays $H^0\rightarrow b\bar b$ and $H^0\rightarrow
Z^0Z^0$.
For the $A^0$, the feasible reactions are $\gamma\gamma\rightarrow
A^0\rightarrow
Z^0h^0/b\bar b$, if $M_{A^0}\Ord 2m_t$, and $\gamma\gamma\rightarrow
A^0\rightarrow t\bar t$,
if $M_{A^0}\OOrd 2m_t$. If $\tan\beta\Ord$ 20, only the $b\bar b$ channel
results
useful for the $A^0$, with $M_{A^0}\Ord 250$ GeV\footnote{Since the $h^0$ mass
never becomes
large, the only important channel is $\gamma\gamma\rightarrow h^0\rightarrow
b\bar b$,
allowing its detection for $M_{h^0}\OOrd 60$ GeV ($M_{A^0}\OOrd 70$ GeV).}.
Recently, it has been shown that the intermediate mass $H^+H^-$ pair production
via $\gamma\gamma$ fusion results greater (e.g., at least of a factor 2 at
$\sqrt s_{ee}=500$ GeV)
than the corresponding $e^+e^-$ mode,
and charged Higgses can be detected using the three decay
modes $\nu_\tau\tau^+\bar\nu_\tau\tau^-$, $c\bar s\bar c s$ and
$c\bar s\bar\nu_\tau\tau^- +\nu_\tau\tau^+ \bar c s$
in a complementary way in order to cover the intermediate mass region of
$H^\pm$ \cite{ggHH}.\par
The $e\gamma$ option at NLCs, obtainable by converting one of the lepton beams
in a photon one using Compton back--scattering, results quite interesting for
the
production of a \sm\ Higgs $\phi$ via the process $e\gamma\rightarrow
\nu_eW\phi$.
For an intermediate Higgs mass this cross section has been demonstrated
\cite{Boos,Hagiwara}
to be comparable to the fusion process
$e^+e^-\rightarrow \bar\nu_e\nu_eW^{\pm*}
W^{\mp*}\rightarrow\bar\nu_e\nu_e\phi$,
and larger than the bremsstrahlung reaction $e^+e^-\rightarrow
Z^{0*}\rightarrow Z^0\phi$,
for $\sqrt s_{ee}$=1--2 TeV.
On the contrary, the similar reaction $e\gamma\rightarrow eZ^0\phi$ (via \sm\
neutral currents)
results smaller by at least an order of magnitude, for corresponding values of
$M_\phi$ and
$\sqrt s_{ee}$. This is due to the $s$--channel behaviour of the associated
matrix element \cite{Boos}.
A detailed analysis of the first process has been carried out in ref.
\cite{Cheung}, where
all possible backgrounds for the signature $W^-\phi\rightarrow jjb\bar b$ have
been
carefully studied, with and without taking into account $b$--tagging.
After appropriate cuts to enhance the signal {\it vs.} background ratio,
one gets that the \sm\ Higgs $\phi$
can be discovered over the mass range 60 GeV $\Ord M_\phi\Ord$ 150 GeV.
Finally, it has been shown in ref. \cite{Eboli} that the process
$e\gamma\rightarrow e\gamma\gamma\rightarrow e\phi$ is the most important
mechanism for $\phi$ production
at $\sqrt s_{ee}=500$ GeV, for $M_\phi\OOrd 140$ GeV.\par
It is the purpose of this paper to study the \mssm\ processes
\begin{equation}
e^-\gamma\rightarrow \nu_eW^-\Phi^0,
\end{equation}
\begin{equation}
e^-\gamma\rightarrow \nu_eH^-\Phi^0,
\end{equation}
\begin{equation}
e^-\gamma\rightarrow e^-Z^0\Phi^0,
\end{equation}
\begin{equation}
e^-\gamma\rightarrow e^-H^+H^-,
\end{equation}
in the intermediate mass range of $M_{A^0}$ and for different values
of $\tan\beta$. Reactions (3) and (5) are the corresponding ones
to $e^-\gamma\rightarrow \nu_e W^-\phi$ and $e^-\gamma \rightarrow e^-
Z^0\phi$.
As in the \sm, although in ref.~\cite{Boos} it is not clear which acceptance
cuts
have been applied to the electron (also in order to avoid singularities from
massless $e$--propagators), process (5) is expected to give contributions
at least an order of magnitude smaller than process (3).\par
The plan of the paper is the following. In section II, we
give some details of the calculation
and the numerical values adopted for the various parameters.
Section III is devoted to a discussion of the results.
Conclusions are in section IV. Finally,
in Appendix, we give the tree--level helicity amplitudes for the
processes (3)--(6) in the formalism of refs. \cite{ks,mana},
and, for convenience of the reader, we collect various formulae which are used
in the analytic expressions.

\subsection*{Calculation}

In the unitary gauge the Feynman diagrams describing at tree--level
the reactions (3)--(6) are shown in figs.~1--4, respectively.
Electron and neutrino are considered massless, so
diagrams with a direct coupling of $\Phi^0/H^\pm$
to the fermion line have been neglected.
For a ${\cal {CP}}$--odd neutral Higgs boson (i.e., $\Phi^0=A^0$), since
diagrams with couplings $VVA^0$ (where $V=W^\pm,Z^0$) vanish,
one has contributions only from process (4).\par
The Feynman amplitudes squared have been computed by the spinor techniques
of refs. \cite{ks,mana}; and,
as a check of the correctness of the results, by the method of ref. \cite{hz},
too.\par
The amplitudes have been  checked for gauge invariance, and
it has been also verified that, with appropriate couplings, our results for
processes
(3) and (5) reproduce those of refs. \cite{Boos,Hagiwara,Cheung}
for a ${\cal {SM}}$ Higgs.\par
Then, matrix elements have been numerically integrated over a
three--body phase space using the Monte Carlo routine VEGAS \cite{vegas}.\par
For the electroweak parameters we have chosen
$s^2_\theta\equiv\sin^2\theta_W=0.23$
and $\alpha_{em}=1/128$, with the masses $M_{Z^0} = 91.175$ GeV and
$M_{W^\pm}=M_{Z^0}c_\theta$ (with $c_\theta\equiv\cos\theta_W$), being
$\theta_W$ the Weinberg angle.\par
We have used the energy spectrum of the back--scattered (unpolarized) photon
given by \cite{gammaf}
\begin{equation}
F_{\gamma/e} (x)= \frac{1}{D(\xi)}\left[1-x+\frac{1}{1-x}-\frac{4x}{\xi(1-x)}
      +\frac{4x^2}{\xi^2(1-x)^2}\right],
\end{equation}
where $D(\xi)$ is the normalization factor
\begin{equation}
D(\xi)=\left(1-\frac{4}{\xi}-\frac{8}{\xi^2}\right)\ln(1+\xi)
+\frac{1}{2}+\frac{8}{\xi}-\frac{1}{2(1+\xi)^2},
\end{equation}
and $\xi=4E_0\omega_0/m_e^2$, where
$\omega_0$ is the incoming laser photon energy
and $E_0$ the (unpolarized) positron one. In eq.~(7)
$x=\omega/E_0$ is the fraction
of the energy of the incident positron carried by the back--scattered photon,
with a maximum value
\begin{equation}
x_{\mathrm {max}}=\frac{\xi}{1+\xi}.
\end{equation}
In order to maximize $\omega$ avoiding $e^+e^-$ pair creation, one takes
$\omega_0$ such that $\xi=2(1+\sqrt 2)$. So, we obtain the typical
values $\xi\simeq 4.8$, $x_{\mathrm {max}}\simeq 0.83$, $D(\xi)\simeq 1.8$,
with
$\omega_0\simeq 1.25(0.63)$ eV for a $\sqrt s_{ee} = 0.5(1)$ TeV $e^+e^-$
collider.\par
In the case of an $e\gamma$ scattering the total cross section $\sigma$ is
obtained by folding
the subprocess cross section $\hat\sigma$ with the photon luminosity
$F_{\gamma/e}$:
\begin{equation}
\sigma(s_{ee})=\int_{x_{\mathrm {min}}}^{x_{\mathrm {max}}}dx
F_{\gamma/e}(x)\hat\sigma(\hat s_{e\gamma}
=xs_{ee}),
\end{equation}
where $\hat s_{e\gamma}$ is the center
of mass (CM) energy at parton ($e\gamma$) level, while
\begin{equation}
x_{\mathrm {min}}=\frac {(M_{\mathrm {final}})^2}{s_{ee}},
\end{equation}
with $M_{\mathrm {final}}$ the sum of the final state particle masses.\par
For the discussion of the results we have assumed a
total integrated luminosity ${\cal L}=20$ fb$^{-1}$ for
an $e^+e^-$ NLC with CM energy ranging from $\sqrt s_{ee}=500$
GeV up to $\sqrt s_{ee}=2$ TeV.\par
We have analyzed the processes (3)--(6) for the mass values
$M_{A^0}=60$, 110 and 160 GeV, with tan$\beta=2$, 15 and
30, adopting for the ${\cal {MSSM}}$
neutral Higgs masses the one--loop expression \cite{corrmssm}:
\begin{eqnarray}\label{m1}
M^{2}_{h^0,H^0}& = & \frac{1}{2}[M^{2}_{A^0} + M_{Z^0}^{2} +
\epsilon/\sin^{2}\beta] \nonumber \\
           &   & \pm \left\{ [ (M^{2}_{A^0} - M^{2}_{Z^0})\cos2\beta +
\epsilon/\sin^{2}\beta]^{2}
                 +(M^{2}_{A^0} + M^{2}_{Z^0})^{2}{\sin}^{2}2\beta
\right\}^{1/2},
\end{eqnarray}
where
\begin{equation}\label{m2}
\epsilon = \frac{3e^{2}}{8\pi^{2} M^{2}_{W^\pm}s^2_\theta} m_{t}^{4} {\mathrm
{ln}}\left( 1 +
\frac{{m}^{2}_{\tilde t}}{m_{t}^{2}} \right),
\end{equation}
with $-e$ the electron charge.
For the $squark$ mass scale we have taken the value ${m}_{\tilde t}=1$ TeV,
while the $top$ mass is $m_t=150$ GeV.
The mixing angle $\alpha$ in the ${\cal {CP}}$--even sector
is defined by
\begin{equation}\label{m3}
\tan 2\alpha = \frac{(M_{A^0}^{2} + M_{Z^0}^{2}){\sin}2\beta}{(M_{A^0}^{2} -
M_{Z^0}^{2})
{\cos2}\beta + \epsilon/{\sin}^{2}\beta}.
\end{equation}
For the ${\cal {MSSM}}$ charged Higgs masses we have maintained the
tree--level expression
\begin{equation}
M_{H^\pm}^2=M_{A^0}^2+M_{W^\pm}^2,
\end{equation}
since one--loop corrections are quite small if compared
with the corresponding ones for neutral Higgses \cite{bri1}.\par
Finally, we have collected in table I the ${\cal {MSSM}}$
couplings which appear, together with the \sm\ ones of
table II, in the processes (3)--(6).

\subsection*{Results}

Our results are given in fig.~5 through fig.~9 and in table III.
In the figures we have labelled in columns
the ``flavors'' of $\Phi^0$ (i.e., $H^0$, $h^0$ and $A^0$)
corresponding to the various reactions:
on the left for process (3), on the right for process (4), and
in the middle of the figures for process (5).\par
The total cross sections of processes (3)--(6) are presented
in figs.~5--7, for three different values of $\tan\beta$ and
$M_{A^0}$, as a function of the $e^+e^-$ CM energy, in the range
500 GeV $\leq \sqrt s_{ee}\leq$ 2 TeV.
Since the cross sections of process (4), for $\Phi^0=A^0$,
and (6) do not depend on $\tan\beta$, identical curves
appear three times in the same figure: they have been drawn in this way
for an easier comparison with the other ones. In fig.~5(6), at $\tan\beta=15$
and 30, the cross sections of process (4), for $h^0$($H^0$)
and $A^0$, coincide within the integration errors.\par
In figs.~8--9 we present the differential distributions
$d\sigma/dp_T(VV)$, of the transverse momentum of the boson pair,
and $d\sigma/dc_\theta^{12}$, of
$c_\theta^{12}=\cos(\theta_{V_1})\cos(\theta_{V_2})$
the product of the cosines of the angles between the
outgoing boson momenta and the incoming $e^-$ direction, respectively,
with $VV\equiv V_1V_2=W^-\Phi^0,H^-\Phi^0,Z^0\Phi^0$ and $H^+H^-$,
at $\sqrt s_{ee}=1$ and 2 TeV, for the parameter
values $M_{A^0}=110$ GeV and $\tan\beta=15$.
The behaviour of the distributions in figs.~8--9
is typical over the parameter space ad for all the CM energies studied.\par
To select electrons in the final state we have imposed the following
cuts: $p_T^e>1$ GeV and $|\cos\theta_e|<$ 0.9.\par
For the observability of a signal, we require a rate
$S\geq6$ events with a significance $S/\sqrt B>4$
for the discovery of an isolated Higgs peak, while
for the case of Higgs peaks overlapping with $W^\pm/Z^0$ peaks
we require $S\geq10$ with $S/\sqrt B>6$, as in ref. \cite{Cheung}.\par
\vskip1.0cm
\centerline{\sl I. The $H^0$ and $h^0$ \mssm\ Higgs bosons.}
\vskip0.5cm
We can see that for process (3) the cross section for the production of one or
both
the ${\cal {CP}}$--even bosons $H^0$ and $h^0$ is large enough
to give a sizeable number of events over the whole region of the
parameter space $(M_{A^0},\tan\beta)$ here considered,
for $\sqrt s_{ee}\OOrd 1$ TeV.
For example, if $M_{A^0}=60$ GeV and $\tan\beta=2$ (so
$M_{H^0}\approx 125$ GeV) we
have that for $H^0$ the cross section grows from $\approx29$ fb at
$\sqrt s_{ee}=1$ TeV up to $\approx77$ fb at $\sqrt s_{ee}=2$ TeV.
In the same energy
range the rate for $h^0$ (with $M_{h^0}\approx 50$ GeV)
varies from $\approx 56$ fb up to $\approx 130$ fb,
allowing a simultaneous search of both the scalars.
The same thing happens when $M_{A^0}=110$ GeV, for all
the three values $\tan\beta=2,15$ and 30, where the $H^0$($h^0$) mass
assumes the values $M_{H^0(h^0)}\approx147$, 119 and
116 GeV (70, 106 and 108 GeV), respectively.
In the other regions of the parameter space only one
between $H^0$ and $h^0$ can be searched for: in particular,
$H^0$ when $M_{A^0}=60$ GeV (with $\tan\beta=15$ and 30), and $h^0$
for $M_{A^0}=160$ GeV (for all $\tan\beta$).\par
The behaviour of these cross sections can be easily understood in terms
of the mass relations (12)--(15) and the couplings of table I.\par
For process (3), the \mssm\ coupling which appears into the
cross sections is, for $H^0$($h^0$), proportional to
$\cos^2(\beta-\alpha)$($\sin^2(\beta-\alpha)$).
Therefore, at fixed $\sqrt s_{ee}$,
the increase of $\sigma(e^-\gamma\rightarrow\nu_eW^-H^0)$
with $\tan\beta$, for $M_{A^0}=60$ GeV,
is contemporaneously due to the coupling, which monotonically
grows from $\approx 0.41$ (for $\tan\beta=2$) to $\approx 0.99$
(for $\tan\beta=30$) and to the phase space,
since $M_{H^0}$ decreases, always
monotonically, from $\approx 125$ GeV (when $\tan\beta=2$) to
$\approx 115$ GeV (when $\tan\beta=30$), being the first effect the
dominant one. The reaction $\sigma(e^-\gamma\rightarrow\nu_eW^-
h^0)$ shows an opposite trend for corresponding values of $(M_{A^0},\tan\beta)$
both in coupling and in phase space,
ranging the first from $\approx 0.58$ to $\approx5.5\times
10^{-3}$ and $M_{h^0}$ from $\approx50$ GeV
to $\approx60$ GeV, resulting in an overall decrease of the cross section.
Conflicting behaviours between coupling and phase space
appear, e.g., for $h^0$ at $M_{A^0}=160$ GeV, when
both coupling and $M_{h^0}$ increase as
$\tan\beta$ become larger, while the corresponding cross section
decreases if $\sqrt s_{ee}=500$ GeV, being in this case the
phase space effect the dominant one.\par
In the same way, all other behaviours
of the remaining cross sections can be explained.\par
The production rates of $H^0$ and $h^0$ via the processes (4) and
(5) are small ($\Ord$ 10 fb) for all the analyzed values of $M_{A^0}$
and $\tan\beta$, and we do not further discuss them.\par
Incidentally, we note that for $\tan\beta=30$ we approximately
recover the \sm\ cross sections
(i.e., the \mssm\ extra couplings are $\approx1$) for
$e^-\gamma\rightarrow \nu_eW^-\phi$
and $e^-\gamma\rightarrow e^-Z^0\phi$,
when $M_{A^0}=60$ GeV in the case of $H^0$
and when $M_{A^0}=160$ GeV in the case of $h^0$,
for values of $M_\phi$ corresponding to
$M_{H^0}\approx M_{h^0}\approx115$ GeV \cite{Boos,Hagiwara,Cheung}.\par
For process (3), we search for the same \sm\ signature,
$jjb\bar b$ plus missing transverse momentum ${p\Dir}_T$,
with the two jets coming from the $W^-$ and
the $b\bar b$ pair from $H^0/h^0$, resorting
to $b$--tagging. The major backgrounds
are the same ones studied in ref. \cite{Cheung} for the \sm\
process  $e^-\gamma\rightarrow \nu_eW^-\phi$, i.e., boson pair production:
\begin{equation}
e^-\gamma \rightarrow \nu_e W^-Z^0,
\end{equation}
\begin{equation}
e^-\gamma \rightarrow e^- W^+W^-,
\end{equation}
\begin{equation}
e^-\gamma \rightarrow e^- Z^0Z^0,
\end{equation}
and heavy quark pair production:
\begin{equation}
e^-\gamma(e^-g) \rightarrow \nu_eb\bar t,
\end{equation}
\begin{equation}
e^-\gamma\rightarrow  e^-t\bar t,
\end{equation}
(with $top$ semileptonically decaying via $t \rightarrow Wb$), where
$g$ represents a gluon, considered as a constituent of the photon
in a $\gamma$--``resolved''  process \cite{Witten}. For the $e^-W^+W^-$
background (17) it is assumed that one is able to pick out only the $jet$ pair
from the more energetic $W^\pm$.
The cross section of these processes
have been computed in refs. \cite{Boos,Hagiwara,Cheung,CheungNP,Jikia}.\par
As we can see comparing our figs.~8 and 9 with figs.~5 and 6c of \cite{Cheung},
respectively, the $p_T(VV)$ and $c_\theta^{12}$ distributions for
process (3) show, in general, the same dependences of the corresponding
\sm\ case: in particular, the two bosons $W^-$ and $H^0/h^0$, on one hand,
have a low probability to appear in configurations with small transverse
momentum,
and, on the other hand, tend to come out in opposite hemispheres.
So, in order to differentiate the signal from the various backgrounds,
we adopt the same acceptance cuts of ref. \cite{Cheung}:
$p_T(VV)>15(30)$ GeV for $\sqrt s_{ee}=1(2)$ TeV, and $c_\theta^{12}<0$.\par
In table III we present, for $\tan\beta=15$, the
cross sections and significance $S_{\Phi^0}/\sqrt B$
(for ${\cal L}=20$ fb$^{-1}$) of process (3), in the range $M_{A^0}=60\div160$
GeV
at $\sqrt s_{ee}=1$(2) TeV, after the cuts in
$p_T(VV)$ and $c_\theta^{12}$, taking into account
the branching ratios of $W\rightarrow jj$ and
$\Phi^0\rightarrow b\bar b$, and assuming a 100\% $b$--tagging.
The decay rates $\Phi^0\rightarrow b\bar b$
have been estimated from ref. \cite{guide} using
the one--loop mass relations (12)--(15), whereas
for the decay $W\rightarrow jj$ we have assumed a branching
fraction of about 70\%.\par
The values of the cross sections
for the various backgrounds have been estimated from ref. \cite{Cheung},
where the following constrains
have been applied: {\it i)} the above combined cuts in $p_T(VV)$ and
$c_{\theta}^{12}$,
{\it ii)} the requirement that $|M_{\Phi^0}-M_{b\bar b}|<5$ GeV for continuum
backgrounds (heavy quark pair production), or the assumption
that a linear fraction of the $Z^0$ peak falls inside
the $\Phi^0$ peak (see eq.~(10) of ref. \cite{Cheung}) for
discrete backgrounds (boson pair production). Finally, {\it iii)}
events with an electron in the final state such
that $E_e>50$ GeV and $|\cos\theta_e|<\cos(0.15)$
have been rejected (central $e^-$ veto).\par
{}From table III, we can argue that, for $\tan\beta=15$, by employing
$b$--tagging,
the detection of both the neutral ${\cal {CP}}$--even \mssm\ Higgses
should be viable in the range 100 GeV $\Ord M_{A^0}\Ord$ 140
GeV and $M_{A^0}\approx70-80$ GeV.
For $M_{A^0}\approx60$ GeV we expect problems from $h^0$ at $\sqrt s_{ee}=1$
TeV,
whereas for $M_{A^0}\OOrd$ 150 GeV only
the $h^0$ can be detected.
If $M_{A^0}\approx100$ GeV, since the background
is greater than the $h^0$ signal, we need the exact absolute normalization of
the $Z^0$ peak, whereas for $M_{A^0}\approx90$ GeV (so
$M_{h^0}\approx M_{Z^0}$), it appears impossible to distinguish
$h^0\rightarrow b\bar b$ from the $Z^0$ peak.
No difficulties appear for $H^0$ when $M_{A^0}\approx90-100$ GeV.\par
So far we have supposed a 100\% acceptance and detection
efficiencies for $jets$ in the final state, the same
for $b$--tagging. Assuming a 25\% overall efficiency, so
$S_{\Phi^0}/\sqrt B$ results halved, we still
obtain a number of events and a significance large enough to cover the above
mentioned
parameter region, except for the cases: $M_{A^0}\approx140$ GeV at
$\sqrt s_{ee}=1$ TeV for $H^0$, and $M_{A^0}\Ord80(\approx60)$ GeV at
$\sqrt s_{ee}=1(2)$ TeV for $h^0$. For $M_{A^0}\approx100$ GeV
the $h^0$ signal can not be separated from backgrounds.\par
Similar conclusions still hold for $\tan\beta=30$, except for
$M_{H^0}(M_{h^0})\OOrd140(\Ord100)$ GeV, where $H^0(h^0)$ is not
visible. If $\tan\beta=2$,
the rate for $\sigma(e^-\gamma\rightarrow\nu_eW^-\Phi^0)$
ranges, for $H^0$, from $\approx1.4(3.1)$ fb to $\approx1.6(3.8)$ fb
(for $M_{A^0}=60\div130$ GeV)
with a maximum of 6.5(14) fb at $M_{A^0}\approx90$ GeV,
and, for $h^0$, from $\approx27(51)$ fb to $\approx37(73)$ fb
(for $M_{A^0}=60\div160$ GeV),
at $\sqrt s_{ee}=1(2)$ TeV. These numbers include the combined cuts
in $p_T(VV)$ and $c_{\theta}^{12}$ and the branching ratios of
$W^\pm\rightarrow jj$
and $\Phi^0\rightarrow b\bar b$. The total fraction of continuum
backgrounds which falls inside the window $|M_{\Phi^0}-M_{b\bar b}|<5$ GeV
is always quite small, both for $H^0$ and for $h^0$, moreover, now
$M_{h^0}\not\approx M_{Z^0}$
(i.e., $M_{h^0}\Ord 80$ GeV for $M_{A^0}\leq 160$ GeV) and
the peaks from discrete backgrounds do not overlap with $h^0\rightarrow
b\bar b$. Therefore, the two Higgs signals are viable: in the case of
$H^0$ for $M_{A^0}\Ord 130$ GeV, and, for $h^0$, over the whole
intermediate mass range of $A^0$.\par

\vskip1.0cm
\centerline{\sl II. The $A^0$ and $H^\pm$ \mssm\ Higgs bosons.}
\vskip0.5cm
We now turn to study the production of the neutral ${\cal {CP}}$--odd
Higgs $A^0$ and the charged $H^\pm$'s.\par
They can be produced through the processes (4) and (6), and
the corresponding cross sections are observable only
if $M_{A^0}=60$ GeV (so $M_{H^\pm}\approx100$ GeV).
We get $\sigma(e^-\gamma\rightarrow \nu_eH^-A^0)\approx3.7$ fb
for process (4) at $\sqrt s_{ee}=2$ TeV,
and $\sigma(e^-\gamma\rightarrow e^-H^+H^-)\approx3.1-3.0$ fb
for process (6) at $\sqrt s_{ee}=900-1000$ GeV\footnote{For this
latter value of $\sqrt s_{ee}$
we have $\sigma(e^-\gamma\rightarrow \nu_eH^-A^0)\approx2.0-2.2$
fb.}. For an integrated luminosity
${\cal L}=20$ fb$^{-1}$, we can rely on a production rate
of $\approx74$ events for $H^-A^0$ at $\sqrt s_{ee}=2$ TeV, and
of $\approx60$ events for  $H^+H^-$ at $\sqrt s_{ee}=1$
TeV\footnote{In the case of the $H^{\pm}$'s, the final ``inclusive''
result, with both the contributions from processes (4) and (6)
at $\sqrt s_{ee}=1$ TeV and with $M_{A^0}=60$ GeV,
is $\approx104$ events, $\approx44$ of them coming from
$e^-\gamma\rightarrow \nu_e H^-A^0$.}.\par
In the parameter space region considered,  an $A^0$ with mass of
60 GeV decays to $b\bar b$ pairs with a branching ratio
$\approx 1$. For a charged Higgs $H^\pm$,
since $M_{H^\pm}<m_t+m_b$ for
$m_t=150$ GeV, the dominant decay modes are
$\nu_\tau\tau^+(\bar\nu_\tau\tau^-)$ and $c\bar s(\bar c s)$
(the channel $W^\pm h^0$ is open for $M_{H^\pm}\OOrd 125$ GeV,
where however the production rate is small), with
$B(H^+\rightarrow \nu_\tau\tau^+)+B(H^+\rightarrow c\bar s)\approx 1$.
So, signatures will be $c \bar s(\bar\nu_\tau\tau^-)b\bar b$ plus
missing transverse momentum ${p\Dir}_T$
in the case of $\nu_eH^-A^0$, and the three combinations
$\nu_\tau\tau^+\bar\nu_\tau\tau^-$, $c\bar s\bar c s$ and
$c\bar s\bar\nu_\tau\tau^- +\nu_\tau\tau^+ \bar c s$, plus an $e^-$,
for $e^-H^+H^-$.\par
Before analyzing the signals {\it vs.} backgrounds
behaviours of these processes, we make some preliminary considerations.\par
First, we recall that the cross sections $e^-\gamma\rightarrow \nu_eH^-A^0$
and  $e^-\gamma\rightarrow e^-H^+H^-$ do not depend on $\tan\beta$ and we
note that their decrease at large $M_{A^0}/M_{H^\pm}$ values
is due to a phase space effect.
Second, after we have applied the cut in $p_T(VV)$, we obtain a slight
reduction
of the signal rates: there remains
$\approx72$ events for $\nu_eH^-A^0$ at $\sqrt s_{ee}=2$ TeV, and
$\approx58$ ones for $e^-H^+H^-$ at $\sqrt s_{ee}=1$ TeV.
Third, in the case of $VV=H^-A^0(H^+H^-)$
the two scalar bosons appear to come out in the same(with
no preference for one of the two) hemisphere(s), so
we cannot utilize the cut $c_\theta^{12}<0$
in order to to reduce continuum backgrounds.
Fourth, for $e^-\gamma\rightarrow e^-H^+H^-$
we search for signatures with an electron in the final state,
so we do not require the central $e^-$ veto.
Fifth, we search for the production of ``two Higgs bosons''
in both processes.\par
For $\nu_eH^-A^0$ the major backgrounds are the same considered above.
In this case we can require that both
$\bar c s/\bar\nu_\tau\tau^-$ and $b\bar b$ pairs (assuming $b$--tagging)
have invariant masses into
the windows $|M_{A^0,H^\pm}-M_{b\bar b,\bar c s/\bar\nu_\tau\tau^-}|<5$ GeV,
with
$M_{A^0}=60$ GeV and $M_{H^\pm}=100$ GeV.\par
These latter constraints totally eliminate the
discrete backgrounds\footnote{For the
naive assumption that the linear fraction of the
$Z^0$ peak in eq.~(10) of \cite{Cheung} falls inside
the Higgs peak.}, since $b\bar b$ pairs
from $e^-W^+W^-$, $\nu_eW^-Z^0$
and $e^-Z^0Z^0$ never reconstruct $M_{A^0}$.
With regard to the continuum backgrounds,
we estimate that the request of picking
out the combination of Higgs bosons $H^-A^0$
reduce them of a factor approximately equal to the one
obtained in ref.~\cite{Cheung} (see Tables I and II therein) in
the case of the \sm\ signal $\nu_e W^-\phi$, since
the $b\bar b$ invariant mass distributions are rather flat, as
shown in fig.~7 of \cite{Cheung}.
This gives a total background of $B\approx82$ events (after the $p_T(VV)$ cut)
to be compared with the signal rate $S_{H^-A^0}\approx14$ events, for
$B(H^+\rightarrow
c\bar s)\approx0.2$: therefore, the significance is only $S_{H^-A^0}/\sqrt
B\approx1.5$, which is too small for a clear detection.
On the contrary, for the leptonic decay $H^-\rightarrow \bar\nu_\tau\tau^-$ we
have a branching ratio
of $\approx0.8$, and taking into account that $B(W^-\rightarrow
\bar\nu_\tau\tau^-)\approx0.1$,
we expect the signature $H^-A^0\rightarrow (\bar\nu_\tau\tau^-)(b\bar b)$ to
be clearly visible, with a final rate $S_{H^-A^0}\approx58$ events
and a significance of $\approx17$.\par
As the $H^+H^-$ decay to $b\bar b$ pair is strongly
suppressed by the Cabibbo mixing,
in the case of $e^-H^+H^-$ the reactions (19)--(20) do not
constitute a background, whereas the discrete ones
are the same studied so far. Here, the
$W^\pm$ peaks are quite far from the $H^\pm$ ones
($M_{H^\pm}-M_{W^\pm}\approx 20$ GeV), so the two channels
$\nu_\tau\tau^+\bar\nu_\tau\tau^-$ and
$jj\bar\nu_\tau\tau^-$+$\nu_\tau\tau^+ jj$
appear  clearly visible. Taking into account the
branching ratios into leptons and quarks,
we can rely on $\approx37$ and 9+9 events for the signatures
$\nu_\tau\tau^+\bar\nu_\tau\tau^-$ and
$jj\bar\nu_\tau\tau^-$+$\nu_\tau\tau^+ jj$,
respectively. In the case of
the $4jet$ final state, although the only
process  whose peaks
partially overlap the $H^\pm$ ones ($e^-\gamma\rightarrow e^-Z^0Z^0$)
has a rather small cross section
($\approx2.7$ fb at $\sqrt s_{ee}=1$ TeV after
the $p_T(VV)$ cut) we have, for $B(H^\pm\rightarrow jj)\approx0.2$,
a final number of events, $\approx2$,
too small to hope to disentangle signal from
background.\par
Finally, all the above conclusions do not change even if we assume
a 25\% overall efficiency\footnote{Probably, it is necessary to
re--analyze the signals for $A^0$ and $H^\pm$ and the corresponding backgrounds
using a different set of cuts. In any case, this subject is left for
future studies.}.

\subsection*{Conclusions}

In this paper we have computed a number of cross sections
at TeV $e\gamma$ colliders
for the production of \mssm\ Higgses, both neutral and charged ones,
in the intermediate mass range of $A^0$ and for different values
of $\tan\beta$, taking into account $b$--tagging.\par
For the expected luminosities,
we have found that a CM energy of at least $\approx1$ TeV is required
for a realistic search through the analyzed channels.
Starting from this value of $\sqrt s_{ee}$ it should be possible the discovery,
over the whole intermediate parameter space, of at least
one between the two ${\cal {CP}}$--even neutral Higgses $H^0$ and $h^0$,
via the channel $jjb\bar b$.
In the case of the ${\cal {CP}}$--odd $A^0$ and the
charged $H^\pm$'s,
the situation is more complicated because of the smallness of the corresponding
production rates, but, if $M_{A^0}\approx 60$ GeV (so $M_{H^\pm}\approx 100$
GeV),
their detection appears feasible, resorting to the leptonic decay
$H^\pm\rightarrow \nu_\tau\tau^+(\bar\nu_\tau\tau^-)$, with
$A^0\rightarrow b\bar b$.\par

\subsection*{Appendix}

In this section we report the explicit formulae for the helicity amplitudes of
the
processes (3)--(6).
For completeness, we briefly recall some expressions and functions
used in the following and extracted from refs. \cite{ks,mana,noiNP}.
\vskip 1.0cm
\centerline{{\sl $S$, $Y$ and $Z$ functions.}}
\vskip 0.5cm
Using the definitions introduced in \cite{ks} and
\cite{mana}, by the relations\footnote{Here, $p$($\lambda$) represent a
generic (anti)spinor quadrimomentum(helicity).}
\begin{equation}
\mu=\pm {m\over{\eta}}, \quad\quad\quad \eta=\sqrt{2(p\cdot k_0)},
\end{equation}
where the sign $+(-)$ refers to $u$ ($v$) spinors of quadrimomentum $p$ (with
$p^2=m^2$),
such that\footnote{From now on, unless stated otherwise, we shall use the
symbol
$u$ for both spinors and antispinors.}
\begin{equation}
\sum_{\lambda=\pm} u(p,\lambda)\bar u(p,\lambda)=
p\Dir\pm m,
\end{equation}
one can compute the functions
\begin{equation}
S(\lambda,p_1,p_2)=[\bar u(p_1,\lambda) u(p_2,-\lambda)],
\end{equation}
\begin{equation}
Y(p_1,\lambda_1;p_2,\lambda_2;c_R,c_L)=
[\bar u(p_1,\lambda_1) \Gamma u(p_2,\lambda_2)],
\end{equation}
and
$$
\hskip -1.2in
Z(p_1,\lambda_1;p_2,\lambda_2;p_3,\lambda_3;p_4,\lambda_4;
c_R,c_L;c'_R,c'_L)=
$$
\begin{equation}
\hskip1.7in
[\bar u(p_1,\lambda_1) \Gamma^{\mu} u(p_2,\lambda_2)]
[\bar u(p_3,\lambda_3) \Gamma'_{\mu} u(p_4,\lambda_4)],
\end{equation}
where
\begin{equation}
\Gamma^{(')\mu}=\gamma^{\mu}\Gamma^{(')},
\end{equation}
and
\begin{equation}
\Gamma^{(')}=c^{(')}_R P_R + c^{(')}_L P_L,
\end{equation}
with
\begin
{equation}P_R={{1+\gamma_5}\over{2}},\quad\quad\quad
P_L={{1-\gamma_5}\over{2}},
\end{equation}
the chiral projectors. \par
Computing the resulting traces one easily finds ($\epsilon^{0123} = 1$)
\cite{ks,mana,noiNP}
\begin{equation}
S(+,p_1,p_2)= 2{{(p_1\cdot k_0)(p_2\cdot k_1)
 -(p_1\cdot k_1)(p_2\cdot k_0)
 +i\epsilon_{\mu\nu\rho\sigma}
  k^\mu_0k^\nu_1p^\rho_1p^\sigma_2}\over{\eta_1\eta_2}},
\end{equation}
$$Y(p_1,+;p_2,+;c_R,c_L)=
c_R\mu_1\eta_2+c_L\mu_2\eta_1,$$
\begin{equation}
Y(p_1,+;p_2,-;c_R,c_L)=c_L S(+,p_1,p_2),
\end{equation}
and
$$\hskip-2.in
Z(p_1,+;p_2,+;p_3,+;p_4,+;c_R,c_L;c'_R,c'_L)=$$
$$\hskip.6in
-2[S(+,p_3,p_1)S(-,p_4,p_2)c'_Rc_R
-\mu_1\mu_2\eta_3\eta_4c'_Rc_L
-\eta_1\eta_2\mu_3\mu_4c'_Lc_R],$$
$$\hskip-2.in
Z(p_1,+;p_2,+;p_3,+;p_4,-;c_R,c_L;c'_R,c'_L)=$$
$$\hskip1.in
-2\eta_2c_R[S(+,p_4,p_1)\mu_3c'_L-S(+,p_3,p_1)\mu_4c'_R],$$
$$\hskip-2.in
Z(p_1,+;p_2,+;p_3,-;p_4,+;c_R,c_L;c'_R,c'_L)=$$
$$\hskip1.in
-2\eta_1c_R[S(-,p_2,p_3)\mu_4c'_L-S(-,p_2,p_4)\mu_3c'_R],$$
$$\hskip-2.in
Z(p_1,+;p_2,+;p_3,-;p_4,-;c_R,c_L;c'_R,c'_L)=$$
$$\hskip.6in
-2[S(+,p_1,p_4)S(-,p_2,p_3)c'_Lc_R
-\mu_1\mu_2\eta_3\eta_4c'_Lc_L
-\eta_1\eta_2\mu_3\mu_4c'_Rc_R],$$
$$\hskip-2.in
Z(p_1,+;p_2,-;p_3,+;p_4,+;c_R,c_L;c'_R,c'_L)=$$
$$\hskip.9in
-2\eta_4c'_R[S(+,p_3,p_1)\mu_2c_R-S(+,p_3,p_2)\mu_1c_L]$$
$$\hskip-1.85in
Z(p_1,+;p_2,-;p_3,+;p_4,-;c_R,c_L;c'_R,c'_L)=0,$$
$$\hskip-2.in
Z(p_1,+;p_2,-;p_3,-;p_4,+;c_R,c_L;c'_R,c'_L)=$$
$$\hskip.6in
-2[\mu_1\mu_4\eta_2\eta_3c'_Lc_L
+\mu_2\mu_3\eta_1\eta_4c'_Rc_R
-\mu_2\mu_4\eta_1\eta_3c'_Lc_R
-\mu_1\mu_3\eta_2\eta_4c'_Rc_L],$$
$$
\hskip-2.in
Z(p_1,+;p_2,-;p_3,-;p_4,-;c_R,c_L;c'_R,c'_L)=$$
\begin{equation}
\hskip1.in
-2\eta_3c'_L[S(+,p_2,p_4)\mu_1c_L-S(+,p_1,p_4)\mu_2c_R].
\end{equation}
\vskip 0.5cm
For the $S$ functions, we have
\begin{equation}
S(-,p_1,p_2)= S(+,p_2,p_1)^*,
\end{equation}
while the remaining $Y$ and $Z$ functions can be obtained by exchanging
$+\leftrightarrow -$ and $R\leftrightarrow L$.
In the previous equations $k_0$  and $k_1$ are two arbitrary
four--vectors (see ref. \cite{ks}) such that
\begin{equation}
k_0\cdot k_0=0, \quad\quad k_1\cdot k_1=-1, \quad\quad k_0\cdot k_1=0.
\end{equation}

Now, we introduce the definitions:
\begin{equation}
-b_1=-b_2=b_3=2b_4=2b_5=2b_6=2b_7=1
\end{equation}
for the coefficients of the incoming/outgoing quadrimomenta,
\begin{equation}
D(p)\equiv D_{\gamma}(p)={1\over {p^2}},
\quad\quad
D_{V}(p)={1\over {p^2-M_V^2}}
\end{equation}
for the propagators, where $V=W^\pm,H^\pm$ or $Z^0$,
\begin{equation}
N_2=[4(p_2\cdot q_2)]^{-1/2}
\end{equation}
for the photon normalization factor,
where $p_2$($q_2$) is the photon quadrimomentum(any four--vector not
proportional
to $p_2$) \cite{ks}.
The symbols $r_1$ and $r_2$ represent two lightlike four--vectors satisfying
the relations
\begin{equation}
r_1^2=r_2^2=0,\quad\quad\quad r_1^\mu+r_2^\mu=p^\mu_4,
\end{equation}
($d\Omega_{r_1(r_2)}$ indicates the solid angle of $r_{1(2)}$ in the rest frame
of $p_4$)
\cite{ks}, $p_6$ and $p_7$ are antispinor quadrimomenta such that
\begin{equation}
p_6^\mu\equiv p_4^\mu, \quad\quad\quad p_7^\mu\equiv p_5^\mu,
\end{equation}
and
\begin{equation}
\sum_{\lambda=\pm} u(p_i,\lambda)\bar u(p_i,\lambda)=
{p\Dir}_i - m_i,\quad\quad{\mathrm {with}}~i=6,7,
\end{equation}
while\footnote{In writing the matrix elements we have tacitly used
for the massive boson quadrimomenta $p_4$ and $p_5$ the equality
\begin{equation}
{p\Dir}_i={1\over 2}\left(\sum_{\lambda=\pm} u(p_i,\lambda)\bar u(p_i,\lambda)+
\sum_{\lambda=\pm} v(p_i,\lambda)\bar v(p_i,\lambda)\right),\quad\quad{\mathrm
{with}}~i=4,5,
\end{equation}
re--labelling the antispinor quadrimomenta: $v(p_i,\lambda)\rightarrow
v(p_{i+2},\lambda)$,
with $i=4,5$.}
\begin{equation}
\sum_{\lambda=\pm} u(p_i,\lambda)\bar u(p_i,\lambda)=
{p\Dir}_i + m_i,\quad\quad{\mathrm {with}}~i=4,5.
\end{equation}
We also define the mass relation
\begin{equation}
\Delta_{M_{\Phi^0}}=\frac{M_{H^\pm}^2-M_{\Phi^0}^2}{M_{W^\pm}^2},
\end{equation}
and the spinor functions\footnote{We adopt the symbol ${\{\lambda\}}$ to denote
a set
of helicities of all external particles in a given reaction,
$\sum_{\{\lambda\}}$ to indicate the usual sum over all their
possible combinations, and the symbol $\sum_{i=j,k;l}$
to indicate a sum over $i$ from $j$ to $k$ with step $l$.}
$$
{\cal X}_2=\sum_{\lambda=\pm}\sum_{i=1,3;2}b_i
Y(p_2,\lambda_2;p_i,\lambda;1,1)
Y(p_i,\lambda;q_2,\lambda_2;1,1),
$$
$$
{\cal X}_4=\sum_{\lambda=\pm}\sum_{i=5,7;2}b_i
Y(r_1,-;p_i,\lambda;1,1)
Y(p_i,\lambda;r_2,-;1,1),
$$
$$
{\cal X}^{V(')}_{31}=
\sum_{\lambda=\pm}\sum_{i=4,6(5,7);2}b_i
Y(p_3,\lambda_3;p_i,\lambda;1,1)
Y(p_i,\lambda;p_1,\lambda_1;c^e_{R_V},c^e_{L_V}),
$$
$$
{\cal Y}^{(')}_2=\sum_{\lambda=\pm}
\sum_{i=4,6(5,7);2}b_i
Y(p_2,\lambda_2;p_i,\lambda;1,1)
Y(p_i,\lambda;q_2,\lambda_2;1,1),
$$
$$
{\cal Y}_4=\sum_{\lambda=\pm}
Y(r_1,-;p_2,\lambda;1,1)
Y(p_2,\lambda;r_2,-;1,1),
$$
$$
{\cal Y}^{V}_{31}=\sum_{\lambda=\pm}
Y(p_3,\lambda_3;p_2,\lambda;1,1)
Y(p_2,\lambda;p_1,\lambda_1;c^e_{R_V},c^e_{L_V}),
$$
$$
{\cal Z}_{24}=Z(p_2,\lambda_2;q_2,\lambda_2;r_1,-;r_2,-;1,1;1,1),
$$
$$
{\cal Z}_{312}^V=Z(p_3,\lambda_3;p_1,\lambda_1;
p_2,\lambda_2;q_2,\lambda_2;c^e_{R_V},c^e_{L_V};1,1),
$$
\begin{equation}
{\cal Z}_{314}^V=Z(p_3,\lambda_3;p_1,\lambda_1;
r_1,-;r_2,-;c^e_{R_V},c^e_{L_V};1,1),
\end{equation}
where $V$ represents a gauge boson $W^\pm$, $Z^0$ or $\gamma$.\par
Finally, the expressions for the couplings ${\cal H}$ ($c_R$ and $c_L$)
are given in table I (II).\par
\vskip 1.0cm
\centerline{\sl 1. The process $e^-\gamma\rightarrow \nu_eW^-\Phi^0$.}
\vskip 0.5cm
The matrix element, summed on final spins and averaged on initial ones,
of the process
\begin{equation}
e^-(p_1,\lambda_1) + \gamma (p_2,\lambda_2)
\longrightarrow \nu_e (p_3,\lambda_3) + W^- (p_4) + \Phi^0 (p_5),
\end{equation}
where $\Phi^0=H^0$ or $h^0$, is given
by
\begin{equation}
{\left|{\overline M}\right|}=
{1\over 4} {\cal C}^2|{\cal H}_{\Phi^0 W^\pm W^\mp}|^2
N_2^2{3\over {8\pi}}
\sum_{\{\lambda\}}
\delta_{\lambda_1-}\delta_{\lambda_3-}
\int d\Omega_{r_1(r_2)}
\sum_{l,m=1}^{3}{T}_l^{\{\lambda\}} T_m^{\{\lambda\}*},
\end{equation}
where ${\cal C}={e^3}/({\sqrt 2 s^2_\theta})$, and
$${\mathrm
i}T_1^{\{\lambda\}}=D_{W^\pm}(p_1-p_3)D_{W^\pm}(p_2-p_4)M_1^{\{\lambda\}},$$
$${\mathrm
i}T_2^{\{\lambda\}}=D_{W^\pm}(p_1-p_3)D_{W^\pm}(p_4+p_5)M_2^{\{\lambda\}},$$
\begin{equation}
{\mathrm i}T_3^{\{\lambda\}}=D_{W^\pm}(p_4+p_5)D(p_1+p_2)M_3^{\{\lambda\}}.
\end{equation}
We have
$$
M_1^{\{\lambda\}}=2({\cal Y}^W_{31}{\cal Z}_{24}-{\cal Y}_{2}{\cal Z}^W_{314}
-{\cal Y}_{4}{\cal Z}^W_{312}),
$$
$$
M_2^{\{\lambda\}}=2({\cal Y}^W_{31}{\cal Z}_{24}+{\cal X}_{2}{\cal Z}^W_{314}
-{\cal Y}_{4}{\cal Z}^W_{312})+(1+\frac{2p_1\cdot p_3}{M_{W^\pm}^{2}}){\cal
X}_{4}{\cal Z}^W_{312},
$$
$$\hskip-1.0cm
M_3^{\{\lambda\}}=\sum_{\lambda=\pm}\sum_{i=1,2;1}[(-b_i)Z(p_3,\lambda_3;p_i,\lambda;r_1,-;r_2,-;
c^e_{R_{W^\pm}},c^e_{L_{W^\pm}};1,1)
$$
$$
\times
Z(p_i,\lambda;p_1,\lambda_1;p_2,\lambda_2;q_2,\lambda_2;c^e_{R_\gamma},c^e_{L_\gamma};1,1)
$$
\begin{equation}
\hskip1.0cm
-2p_1\cdot p_2{{{\cal
X}_4}\over{M_{W^\pm}^2}}(p_3,\lambda_3;p_i,\lambda;p_2,\lambda_2;q_2,\lambda_2;
c^e_{L_{W^\pm}},c^e_{R_{W^\pm}};1,1)].
\end{equation}
\vskip 1.0cm
\centerline{\sl 2. The process $e^-\gamma\rightarrow \nu_eH^-\Phi^0$.}
\vskip 0.5cm
The Feynman amplitude squared of the reaction
\begin{equation}
e^-(p_1,\lambda_1) + \gamma (p_2,\lambda_2)
\longrightarrow \nu_e (p_3,\lambda_3) + H^- (p_4) + \Phi^0 (p_5),
\end{equation}
where $\Phi^0=H^0,h^0$ or $A^0$, is
\begin{equation}
{\left|{\overline M}\right|}=
{1\over 4} {\cal D}^2 |{\cal H}_{\Phi^0 W^\pm H^\mp (\gamma)}|^2
N_2^2 \sum_{\{\lambda\}}
\delta_{\lambda_1-}\delta_{\lambda_3-}
\sum_{l,m=1}^{4}{T}_l^{\{\lambda\}} T_m^{\{\lambda\}*},
\end{equation}
where ${\cal D}={\cal C}/2$, and
$$-{\mathrm
i}T_1^{\{\lambda\}}=D_{W^\pm}(p_1-p_3)D_{H^\pm}(p_2-p_4)M_1^{\{\lambda\}},$$
$$-{\mathrm
i}T_2^{\{\lambda\}}=D_{W^\pm}(p_1-p_3)D_{W^\pm}(p_4+p_5)M_2^{\{\lambda\}},$$
$$-{\mathrm i}T_3^{\{\lambda\}}=D_{W^\pm}(p_1-p_3)M_3^{\{\lambda\}},$$
\begin{equation}
-{\mathrm i}T_4^{\{\lambda\}}=D_{W^\pm}(p_4+p_5)D(p_1+p_2)M_4^{\{\lambda\}}.
\end{equation}
In this case, one gets
$$
M_1^{\{\lambda\}}=4{\cal Y}_2({\cal Y}^W_{31}-{\cal X}^W_{31}),
$$
$$\hskip -2.7cm
M_2^{\{\lambda\}}=2{\cal Y}^W_{31}[2{\cal Y}_{2}+(1+\Delta_{M_{\Phi^0}}){\cal
X}_2]
$$
$$
+2{\cal X}_2[2{\cal X}^{W}_{31}-(1+\Delta_{M_{\Phi^0}}){\cal Y}^W_{31}]
$$
$$\hskip 5.0cm
+{\cal
Z}^W_{312}(p_4+p_5-2p_2)\cdot[(1-\Delta_{M_{\Phi^0}})p_4-(1+\Delta_{M_{\Phi^0}})p_5],
$$
$$
M_3^{\{\lambda\}}={\cal Z}^W_{312},
$$
$$\hskip -3.0cm
M_4^{\{\lambda\}}=\sum_{\lambda=\pm}\sum_{i=1,2;1}(-b_i)Z(p_i,\lambda;p_1,\lambda_1;
p_2,\lambda_2;q_2,\lambda_2;c^e_{R_\gamma},c^e_{L_\gamma};1,1)
$$
$$\hskip -1.0cm \times
[(1-\Delta_{M_{\Phi^0}})\sum_{\lambda'=\pm}\sum_{j=4,6;2}b_j
Y(p_3,\lambda_3;p_j,\lambda';1,1)Y(p_j,\lambda';p_i,\lambda;c^e_{R_{W^\pm}},c^e_{L_{W^\pm}})
$$
\begin{equation}
\hskip -0.1cm
-(1+\Delta_{M_{\Phi^0}})\sum_{\lambda'=\pm}\sum_{j=5,7;2}b_j
Y(p_3,\lambda_3;p_j,\lambda';1,1)Y(p_j,\lambda';p_i,\lambda;c^e_{R_{W^\pm}},c^e_{L_{W^\pm}})].
\end{equation}
\vskip 1.0cm
\centerline{\sl 3. The process $e^-\gamma\rightarrow e^-Z^0\Phi^0$.}
\vskip 0.5cm
For the process
\begin{equation}
e^-(p_1,\lambda_1) + \gamma (p_2,\lambda_2)
\longrightarrow e^- (p_3,\lambda_3) + Z^0 (p_4) + \Phi^0 (p_5),
\end{equation}
where $\Phi^0=H^0$ or $h^0$, we have the formula
\begin{equation}
{\left|{\overline M}\right|}=
{1\over 4} {\cal E}^2 |{\cal H}_{\Phi^0 Z^0 Z^0}|^2
N_2^2{3\over {8\pi}}
\sum_{\{\lambda\}}
\delta_{\lambda_1\lambda_3}
\int d\Omega_{r_1(r_2)}
\sum_{l,m=1}^{2}{T}_l^{\{\lambda\}} T_m^{\{\lambda\}*},
\end{equation}
where ${\cal E}={e^3}/({s^2_\theta c^2_\theta})$, and
$$
{\mathrm i}T_1^{\{\lambda\}}=D_{Z^0}(p_4+p_5)D(p_1+p_2)M_1^{\{\lambda\}},
$$
\begin{equation}
{\mathrm i}T_2^{\{\lambda\}}=D_{Z^0}(p_4+p_5)D(p_2-p_3)M_2^{\{\lambda\}},
\end{equation}
with
$$\hskip-1.0cm
M_1^{\{\lambda\}}=\sum_{\lambda=\pm}\sum_{i=1,2;1}[(-b_i)Z(p_3,\lambda_3;p_i,\lambda;r_1,-;r_2,-;
c^e_{R_{Z^0}},c^e_{L_{Z^0}};1,1)
$$
$$
\times
Z(p_i,\lambda;p_1,\lambda_1;p_2,\lambda_2;q_2,\lambda_2;c^e_{R_\gamma},c^e_{L_\gamma};1,1)
$$
$$\hskip1.0cm
-2p_1\cdot p_2{{{\cal
X}_4}\over{M_{Z^0}^2}}(p_3,\lambda_3;p_i,\lambda;p_2,\lambda_2;q_2,\lambda_2;
c^e_{L_{Z^0}},c^e_{R_{Z^0}};1,1)],
$$
$$\hskip-1.0cm
M_2^{\{\lambda\}}=\sum_{\lambda=\pm}\sum_{i=2,3;1}[b_iZ(p_3,\lambda_3;p_i,\lambda;p_2,\lambda_2;q_2,\lambda_2;
c^e_{R_\gamma},c^e_{L_\gamma};1,1)
$$
$$
\times Z(p_i,\lambda;p_1,\lambda_1;r_1,-;r_2,-;c^e_{R_{Z^0}},c^e_{L_{Z^0}};1,1)
$$
\begin{equation}
\hskip1.0cm
-2p_2\cdot p_3{{{\cal
X}_4}\over{M_{Z^0}^2}}(p_3,\lambda_3;p_i,\lambda;p_2,\lambda_2;q_2,\lambda_2;
c^e_{L_{Z^0}},c^e_{R_{Z^0}};1,1)].
\end{equation}
\vskip 1.0cm
\centerline{\sl 4. The process $e^-\gamma\rightarrow e^- H^+H^-$.}
\vskip 0.5cm
The Feynman amplitude squared of
\begin{equation}
e^-(p_1,\lambda_1) + \gamma (p_2,\lambda_2)
\longrightarrow e^- (p_3,\lambda_3) + H^- (p_4) + H^+ (p_5)
\end{equation}
is given by the formula
\begin{equation}
{\left|{\overline M}\right|}=
{1\over 4} {\cal F}^2N_2^2
\sum_{\{\lambda\}}
\delta_{\lambda_1\lambda_3}
\sum_{l,m=1}^{5}{T}_l^{\{\lambda\}} T_m^{\{\lambda\}*},
\end{equation}
where ${\cal F}=e^3$, and
\begin{equation}
{T}_l^{\{\lambda\}}={T}_{l,\gamma}^{\{\lambda\}}
+\frac{\cos(2\theta_W)}{2s^2_\theta c^2_\theta}{T}_{l,Z^0}^{\{\lambda\}},
\end{equation}
with ($V=\gamma$ or $Z^0$)
$$-{\mathrm
i}T_{1,V}^{\{\lambda\}}=D_V(p_1-p_3)D_{H^\pm}(p_2-p_5)M_{1,V}^{\{\lambda\}},$$
$$-{\mathrm
i}T_{2,V}^{\{\lambda\}}=D_V(p_1-p_3)D_{H^\pm}(p_2-p_4)M_{2,V}^{\{\lambda\}},$$
$$-{\mathrm i}T_{3,V}^{\{\lambda\}}=-2D_V(p_1-p_3)M_{3,V}^{\{\lambda\}},$$
$$-{\mathrm
i}T_{4,V}^{\{\lambda\}}=D_V(p_4+p_5)D(p_1+p_2)M_{4,V}^{\{\lambda\}},$$
\begin{equation}
-{\mathrm i}T_{5,V}^{\{\lambda\}}=D_V(p_4+p_5)D(p_2-p_3)M_{5,V}^{\{\lambda\}}.
\end{equation}
For the $M_{l,V}^{\{\lambda\}}$'s one has
$$
M_{1,V}^{\{\lambda\}}=-4{\cal Y}'_2({\cal Y}^V_{31}-{\cal X}^{V'}_{31}),
$$
$$
M_{2,V}^{\{\lambda\}}=-4{\cal Y}_2({\cal Y}^V_{31}-{\cal X}^V_{31}),
$$
$$
M_{3,V}^{\{\lambda\}}={\cal Z}^V_{312},
$$
$$\hskip-1.0cm
M_{4,V}^{\{\lambda\}}=\sum_{\lambda=\pm}\sum_{i=1,2;1}(-b_i)
Z(p_i,\lambda;p_1,\lambda_1;p_2,\lambda_2;q_2,\lambda_2;c^e_{R_\gamma},c^e_{L_\gamma};1,1)
$$
$$\times
[\sum_{\lambda'=\pm}\sum_{j=4,6;2}b_j
Y(p_3,\lambda_3;p_j,\lambda';1,1)Y(p_j,\lambda';p_i,\lambda;c^e_{R_V},c^e_{L_V})
$$
$$
-\sum_{\lambda'=\pm}\sum_{j=5,7;2}b_j
Y(p_3,\lambda_3;p_j,\lambda';1,1)Y(p_j,\lambda';p_i,\lambda;c^e_{R_V},c^e_{L_V})],
$$
$$\hskip-1.0cm
M_{5,V}^{\{\lambda\}}=\sum_{\lambda=\pm}\sum_{i=2,3;1}b_iZ(p_3,\lambda_3;p_i,\lambda;
p_2,\lambda_2;q_2,\lambda_2;c^e_{R_\gamma},c^e_{L_\gamma};1,1)
$$
$$\times
[\sum_{\lambda'=\pm}\sum_{j=4,6;2}b_j
Y(p_i,\lambda;p_j,\lambda';1,1)Y(p_j,\lambda';p_1,\lambda_1;c^e_{R_V},c^e_{L_V})
$$
\begin{equation}
-\sum_{\lambda'=\pm}\sum_{j=5,7;2}b_j
Y(p_i,\lambda;p_j,\lambda';1,1)Y(p_j,\lambda';p_1,\lambda_1;c^e_{R_V},c^e_{L_V})].
\end{equation}

\subsection*{Acknowledgements}

We are grateful to R.~Tateo for the interesting discussions, and to
A.~Ballestrero and E.~Maina for a careful reading of the manuscript
and useful suggestions.

\newpage
\subsection*{Table Captions}
\begin{description}
\item[table I  ] ${\cal {MSSM}}$ neutral Higgs couplings
to the gauge bosons $W^\pm$, $Z^0$ and $\gamma$,
and to the charged Higgses $H^\pm$'s.
\item[table II ] Right and left handed couplings
of the leptons $f=e,\nu_e$ to the gauge bosons $W^\pm$, $Z^0$ and $\gamma$.
We have $(Q^e, T^e_3)=(-1, -{1\over 2})$ and
$(Q^{\nu_e}, T^{\nu_e}_3)=(0, {1\over 2})$.
\item[table III] Cross sections and significance $S_{\Phi^0}/\sqrt B$
(for ${\cal L}=20$ fb$^{-1}$) of process (3)
for $H^0/h^0$, with $\tan\beta=15$, in the range $M_{A^0}=60\div160$ GeV
at $\sqrt s_{ee}=1$(2) TeV, after the cuts in
$p_T(VV)$ and $c_\theta^{12}<0$, taking into account
the branching ratios of $W\rightarrow jj$ and
$\Phi^0\rightarrow b\bar b$, and assuming 100\% $b$--tagging.

\end{description}

\vspace*{\fill}

\subsection*{Figure Captions}
\begin{description}
\item[fig.~1 ] Feynman diagrams contributing in the lowest order to
$e^-\gamma\rightarrow \nu_eW^-\Phi^0$, where $\Phi^0=H^0$ or $h^0$,
in the unitary gauge.
Internal wavy lines represent a $W^\pm$.
\item[fig.~2 ] Feynman diagrams contributing in the lowest order to
$e^-\gamma\rightarrow \nu_eH^-\Phi^0$, where $\Phi^0=H^0,h^0$ or $A^0$,
in the unitary gauge.
Internal wavy (dashed) lines represent a $W^\pm$ ($H^\pm$).
\item[fig.~3 ] Feynman diagrams contributing in the lowest order to
$e^-\gamma\rightarrow e^-Z^0\Phi^0$, where $\Phi^0=H^0$ or $h^0$,
in the unitary gauge.
Internal wavy lines represent a $Z^0$.
\item[fig.~4 ] Feynman diagrams contributing in the lowest order to
$e^-\gamma\rightarrow e^-H^+H^-$, in the unitary gauge.
Internal wavy (dashed) lines represent a $Z^0$,$\gamma$ ($H^\pm$).
\item[fig.~5 ] Total cross sections in fb of the processes (3)--(6)
{\it vs.} the $e^+e^-$ CM energy for $\tan\beta=2,15$ and 30, with
$M_{A^0}=60$ GeV. For $\tan\beta=15$ and 30
the cross sections of process (4), for $h^0$
and $A^0$, coincide within the integration errors.
\item[fig.~6 ] Total cross sections in fb of the processes (3)--(6)
{\it vs.}  the $e^+e^-$ CM energy for $\tan\beta=2,15$ and 30, with
$M_{A^0}=110$ GeV.
\item[fig.~7 ] Total cross sections in fb of the processes (3)--(6)
{\it vs.}  the $e^+e^-$ CM energy for $\tan\beta=2,15$ and 30, with
$M_{A^0}=160$ GeV. For $\tan\beta=15$ and 30
the cross sections of process (4), for $H^0$
and $A^0$, coincide within the integration errors.
\item[fig.~8 ] Differential distributions $d\sigma/dp_T(VV)$ of
the transverse momentum of the boson pair at $\sqrt s_{ee}=1,2$ TeV,
for $\tan\beta=15$ and $M_{A^0}=110$ GeV.
\item[fig.~9 ] Differential distributions $d\sigma/c_\theta^{12}$ of
$c_\theta^{12}=\cos(\theta_{V_1})\cos(\theta_{V_2})$
the product of the cosines of the angles between the
outgoing boson momenta and the incoming $e^-$ direction,
at $\sqrt s_{ee}=1,2$ TeV, for $\tan\beta=15$ and $M_{A^0}=110$ GeV.
\end{description}

\vspace*{\fill}

\newpage
\pagestyle{empty}
\begin{table}
\begin{center}
\begin{tabular}{|c|c|c|c|}     \hline
\rule[-0.6cm]{0cm}{1.3cm}
$\;\;\;\;\;$  & $H^0$ & $h^0$ & $A^0$  \\ \hline
\rule[-0.6cm]{0cm}{1.3cm}
$W^\pm W^\mp$ & $\cos (\beta - \alpha)$ & $\sin(\beta - \alpha)$ & $0$   \\
\rule[-0.6cm]{0cm}{1.3cm}
$W^{\pm} H^\mp(\gamma)$ & $\sin(\beta - \alpha)$ & $-\cos(\beta - \alpha)$ &
${\mathrm i}$    \\ 
\rule[-0.6cm]{0cm}{1.3cm}
$Z^0Z^0$ & $\cos(\beta - \alpha)$ & $\sin(\beta - \alpha)$ & $0$    \\ \hline
\end{tabular}
\end{center}
\centerline{\Large Table I}
\end{table}
\
\vskip 3.0cm
\
$$\vbox{\tabskip=0pt \offinterlineskip
\halign to240pt{\strut#& \vrule#\tabskip=1.em plus2.0em& \hfil#&
\vrule#& \hfil#&
\vrule#& \hfil#&
\vrule#& \hfil#&
\vrule#\tabskip=0pt\cr \noalign{\hrule}
&  && && && && \cr
&  && &$~\gamma~~$& &$Z\hskip 1.1truecm$& &$~W^\pm~$& \cr
&  && && && && \cr
\noalign{\hrule}
&  && && && && \cr
&  &$c^{f}_{R_V}$& &$e^{f}~$& &$-Q^f s^2_\theta\quad$& &$0~~~$& \cr
&  && && && && \cr
&  &$c^{f}_{L_V}$& &$e^{f}~$& &$T^f_3-Q^f s^2_\theta$& &$1~~~$& \cr
&  && && && && \cr
\noalign{\hrule}
\noalign{\smallskip}\cr}}$$
\centerline{\Large Table II}
\vfill
\newpage
\
\pagestyle{empty}
\begin{table}[htbp]
\begin{center}
\begin{tabular}{|c||c|c|c|c|c|c|}     \hline
\rule[-0.3cm]{0cm}{1.3cm}
$M_{A^0}~{\mathrm {(GeV)}}$ &
$M_{H^0}~{\mathrm {(GeV)}}$ &
$M_{h^0}~{\mathrm {(GeV)}}$ &
$\sigma_{H^0}~{\mathrm {(fb)}}$  &
$\sigma_{h^0}~{\mathrm {(fb)}}$  &
$S_{H^0}/\sqrt B$ &
$S_{h^0}/\sqrt B$  \\ \hline
\rule[-0.3cm]{0cm}{1.3cm}
$60$ & $115$ & $60$ & $31(66)$ & $0.98(1.9)$ & $126(269)$ & $3.5(7.0)$   \\
\rule[-0.3cm]{0cm}{1.3cm}
$70$ & $115$ & $69$ & $32(68)$ & $1.2(2.3)$ & $131(278)$ & $4.2(8.4)$   \\
\rule[-0.3cm]{0cm}{1.3cm}
$80$ & $116$ & $79$ & $32(68)$ & $1.7(3.3)$ & $131(278)$ & $6.0(12)$   \\
\rule[-0.3cm]{0cm}{1.3cm}
$90$ & $116$ & $89$ & $31(67)$ & $2.7(5.4)$ & $127(273)$ & $2.6(3.5)$   \\
\rule[-0.3cm]{0cm}{1.3cm}
$100$ & $117$ & $98$ & $29(62)$ & $5.3(11)$ & $118(253)$ & $7.6(11)$  \\
\rule[-0.3cm]{0cm}{1.3cm}
$110$ & $119$ & $106$ & $21(45)$ & $13(27)$ & $86(184)$ & $51(106)$  \\
\rule[-0.3cm]{0cm}{1.3cm}
$120$ & $124$ & $110$ & $8.7(19)$ & $26(54)$ & $36(78)$ & $102(220)$  \\
\rule[-0.3cm]{0cm}{1.3cm}
$130$ & $132$ & $112$ & $2.9(6.4)$ & $32(67)$ & $12(27)$ & $126(274)$  \\
\rule[-0.3cm]{0cm}{1.3cm}
$140$ & $142$ & $113$ & $1.2(2.7)$ & $33(70)$ & $5.1(12)$ & $129(286)$  \\
\rule[-0.3cm]{0cm}{1.3cm}
$150$ & $151$ & $113$ & $0.57(1.3)$ & $34(71)$ & $2.5(5.8)$ & $133(290)$  \\
\rule[-0.3cm]{0cm}{1.3cm}
$160$ & $161$ & $113$ & $0.32(0.75)$ & $34(71)$ & $1.4(3.4)$ & $133(290)$  \\
\hline
\end{tabular}
\end{center}
\centerline{\Large Table III}
\end{table}
\vfill
\newpage
\pagestyle{empty}
\
\vskip1.0cm
\begin{picture}(10000,8000)
\THICKLINES
\bigphotons
\drawline\fermion[\W\REG](14250,10000)[5000]
\drawarrow[\E\ATBASE](\pmidx,\pmidy)
\drawline\fermion[\W\REG](\pbackx,\pbacky)[5000]
\drawarrow[\E\ATBASE](\pmidx,\pmidy)
\drawline\photon[\S\REG](\pfrontx,\pfronty)[9]
\seglength=1416  \gaplength=300  
\drawline\scalar[\E\REG](9400,5500)[3]
\drawline\photon[\E\REG](\photonbackx,\photonbacky)[5]
\drawline\photon[\W\REG](\photonfrontx,\photonfronty)[5]
\put(14750,9750){$\nu_e$}
\put(2500,9750){$e^-$}
\put(14650,5150){$\Phi^0$}
\put(14750,750){$W^-$}
\put(2500,750){$~\gamma$}
\put(8650,-2000){$(1)$}
\THICKLINES
\bigphotons
\drawline\fermion[\W\REG](34250,10000)[5000]
\drawarrow[\E\ATBASE](\pmidx,\pmidy)
\drawline\fermion[\W\REG](\pbackx,\pbacky)[5000]
\drawarrow[\E\ATBASE](\pmidx,\pmidy)
\drawline\photon[\S\REG](\pfrontx,\pfronty)[9]
\drawline\photon[\E\REG](\photonbackx,\photonbacky)[5]
\seglength=1416  \gaplength=300  
\drawline\scalar[\NE\REG](31750,1300)[2]
\drawline\photon[\W\REG](29250,1000)[5]
\put(34750,9750){$\nu_e$}
\put(22500,9750){$e^-$}
\put(34400,3750){$\Phi^0$}
\put(34750,750){$W^-$}
\put(22500,750){$~\gamma$}
\put(28650,-2000){$(2)$}
\end{picture}
\vskip 3.0cm
\begin{picture}(10000,8000)
\THICKLINES
\bigphotons
\drawline\fermion[\SW\REG](27000,10000)[6000]
\drawarrow[\NE\ATBASE](\pmidx,\pmidy)
\drawline\photon[\SE\REG](\pbackx,\pbacky)[6]
\seglength=1416  \gaplength=300  
\drawline\scalar[\E\REG](\pmidx,\pmidy)[2]
\drawline\fermion[\W\REG](\fermionbackx,\fermionbacky)[6000]
\drawarrow[\E\ATBASE](\pmidx,\pmidy)
\drawline\fermion[\NW\REG](\pbackx,\pbacky)[6000]
\drawarrow[\SE\ATBASE](\pmidx,\pmidy)
\drawline\photon[\SW\REG](\pfrontx,\pfronty)[6]
\put(27500,9750){$\nu_e$}
\put(11000,9750){$e^-$}
\put(28350,3500){$\Phi^0$}
\put(27250,1250){$W^-$}
\put(11250,1500){$~\gamma$}
\put(19300,-1000){$(3)$}
\end{picture}
\vskip 3.0cm
\centerline{\bf\Large Fig.1}
\vfill
\newpage
\pagestyle{empty}
\
\vskip1.0cm
\begin{picture}(10000,8000)
\THICKLINES
\bigphotons
\drawline\fermion[\W\REG](14250,10000)[5000]
\drawarrow[\E\ATBASE](\pmidx,\pmidy)
\drawline\fermion[\W\REG](\pbackx,\pbacky)[5000]
\drawarrow[\E\ATBASE](\pmidx,\pmidy)
\drawline\photon[\S\REG](\pfrontx,\pfronty)[5]
\seglength=1416  \gaplength=300  
\drawline\scalar[\E\REG](\pbackx,\pbacky)[3]
\seglength=1416  \gaplength=300  
\drawline\scalar[\S\REG](\pfrontx,\pfronty)[2]
\seglength=1416  \gaplength=300  
\drawline\scalar[\E\REG](\pbackx,\pbacky)[3]
\drawline\photon[\W\REG](\pfrontx,\pfronty)[5]
\put(14750,9750){$\nu_e$}
\put(2500,9750){$e^-$}
\put(14650,4650){$\Phi^0$}
\put(14750,1450){$H^-$}
\put(2500,1750){$~\gamma$}
\put(8650,-2000){$(1)$}
\THICKLINES
\bigphotons
\drawline\fermion[\W\REG](34250,10000)[5000]
\drawarrow[\E\ATBASE](\pmidx,\pmidy)
\drawline\fermion[\W\REG](\pbackx,\pbacky)[5000]
\drawarrow[\E\ATBASE](\pmidx,\pmidy)
\drawline\photon[\S\REG](\pfrontx,\pfronty)[9]
\drawline\photon[\E\REG](\photonbackx,\photonbacky)[2]
\seglength=1416  \gaplength=300  
\drawline\scalar[\NE\REG](\pbackx,\pbacky)[2]
\seglength=1416  \gaplength=300  
\drawline\scalar[\E\REG](\pfrontx,\pfronty)[2]
\drawline\photon[\W\REG](29250,1000)[5]
\put(34750,9750){$\nu_e$}
\put(22500,9750){$e^-$}
\put(34000,3500){$\Phi^0$}
\put(34750,750){$H^-$}
\put(22500,750){$~\gamma$}
\put(28650,-2000){$(2)$}
\end{picture}
\vskip 3.0cm
\begin{picture}(10000,8000)
\THICKLINES
\bigphotons
\drawline\fermion[\W\REG](14250,10000)[5000]
\drawarrow[\E\ATBASE](\pmidx,\pmidy)
\drawline\fermion[\W\REG](\pbackx,\pbacky)[5000]
\drawarrow[\E\ATBASE](\pmidx,\pmidy)
\drawline\photon[\S\REG](\pfrontx,\pfronty)[9]
\seglength=1416  \gaplength=300  
\drawline\scalar[\NE\REG](\pbackx,\pbacky)[4]
\seglength=1416  \gaplength=300  
\drawline\scalar[\E\REG](\pfrontx,\pfronty)[3]
\drawline\photon[\W\REG](9250,1000)[5]
\put(14750,9750){$\nu_e$}
\put(2500,9750){$e^-$}
\put(14500,5500){$\Phi^0$}
\put(14750,750){$H^-$}
\put(2500,750){$~\gamma$}
\put(8650,-2000){$(3)$}
\THICKLINES
\bigphotons
\drawline\fermion[\SW\REG](36350,10000)[6000]
\drawarrow[\NE\ATBASE](\pmidx,\pmidy)
\drawline\photon[\SE\REG](\pbackx,\pbacky)[3]
\seglength=1416  \gaplength=300  
\drawline\scalar[\E\REG](\pbackx,\pbacky)[2]
\seglength=1416  \gaplength=300  
\drawline\scalar[\SE\REG](\pfrontx,\pfronty)[2]
\drawline\fermion[\W\REG](\fermionbackx,\fermionbacky)[6000]
\drawarrow[\E\ATBASE](\pmidx,\pmidy)
\drawline\fermion[\NW\REG](\pbackx,\pbacky)[6000]
\drawarrow[\SE\ATBASE](\pmidx,\pmidy)
\drawline\photon[\SW\REG](\pfrontx,\pfronty)[6]
\put(36850,9750){$\nu_e$}
\put(20350,9750){$e^-$}
\put(37700,3500){$\Phi^0$}
\put(36600,1250){$H^-$}
\put(20600,1500){$~\gamma$}
\put(28650,-1000){$(4)$}
\end{picture}
\vskip 3.0cm
\centerline{\bf\Large Fig.2}
\vfill
\newpage
\pagestyle{empty}
\
\vskip1.0cm
\begin{picture}(10000,8000)
\THICKLINES
\bigphotons
\drawline\fermion[\SW\REG](16350,10000)[6000]
\drawarrow[\NE\ATBASE](\pmidx,\pmidy)
\drawline\photon[\SE\REG](\pbackx,\pbacky)[6]
\seglength=1416  \gaplength=300  
\drawline\scalar[\E\REG](\pmidx,\pmidy)[2]
\drawline\fermion[\W\REG](\fermionbackx,\fermionbacky)[6000]
\drawarrow[\E\ATBASE](\pmidx,\pmidy)
\drawline\fermion[\NW\REG](\pbackx,\pbacky)[6000]
\drawarrow[\SE\ATBASE](\pmidx,\pmidy)
\drawline\photon[\SW\REG](\pfrontx,\pfronty)[6]
\put(16850,9750){$e^-$}
\put(350,9750){$e^-$}
\put(17700,3500){$\Phi^0$}
\put(16600,1250){$Z^0$}
\put(600,1500){$~\gamma$}
\put(8650,-1000){$(1)$}
\THICKLINES
\bigphotons
\drawline\fermion[\SW\REG](38350,10000)[6000]
\drawarrow[\NE\ATBASE](\pmidx,\pmidy)
\drawline\photon[\SW\REG](\pbackx,\pbacky)[16]
\drawline\fermion[\W\REG](\fermionbackx,\fermionbacky)[6000]
\drawarrow[\E\ATBASE](\pmidx,\pmidy)
\drawline\fermion[\NW\REG](\pbackx,\pbacky)[6000]
\drawarrow[\SE\ATBASE](\pmidx,\pmidy)
\drawline\photon[\SE\REG](\pfrontx,\pfronty)[16]
\seglength=1416  \gaplength=300  
\drawline\scalar[\E\REG](\pmidx,\pmidy)[3]
\put(38850,9750){$e^-$}
\put(22350,9750){$e^-$}
\put(38500,250){$\Phi^0$}
\put(38600,-4750){$Z^0$}
\put(22600,-4500){$~\gamma$}
\put(30650,-7000){$(2)$}
\end{picture}
\vskip 3.0cm
\centerline{\bf\Large Fig.3}
\vfill
\newpage
\pagestyle{empty}
\
\vskip1.0cm
\begin{picture}(10000,8000)
\THICKLINES
\bigphotons
\drawline\fermion[\W\REG](14250,10000)[5000]
\drawarrow[\E\ATBASE](\pmidx,\pmidy)
\drawline\fermion[\W\REG](\pbackx,\pbacky)[5000]
\drawarrow[\E\ATBASE](\pmidx,\pmidy)
\drawline\photon[\S\REG](\pfrontx,\pfronty)[5]
\seglength=1416  \gaplength=300  
\drawline\scalar[\E\REG](\pbackx,\pbacky)[3]
\seglength=1416  \gaplength=300  
\drawline\scalar[\S\REG](\pfrontx,\pfronty)[2]
\seglength=1416  \gaplength=300  
\drawline\scalar[\E\REG](\pbackx,\pbacky)[3]
\drawline\photon[\W\REG](\pfrontx,\pfronty)[5]
\put(14750,9750){$e^-$}
\put(2500,9750){$e^-$}
\put(14650,4650){$H^-$}
\put(14750,1450){$H^+$}
\put(2500,1750){$~\gamma$}
\put(8650,-2000){$(1)$}
\THICKLINES
\bigphotons
\drawline\fermion[\W\REG](34250,10000)[5000]
\drawarrow[\E\ATBASE](\pmidx,\pmidy)
\drawline\fermion[\W\REG](\pbackx,\pbacky)[5000]
\drawarrow[\E\ATBASE](\pmidx,\pmidy)
\drawline\photon[\S\REG](\pfrontx,\pfronty)[5]
\seglength=1416  \gaplength=300  
\drawline\scalar[\E\REG](\pbackx,\pbacky)[3]
\seglength=1416  \gaplength=300  
\drawline\scalar[\S\REG](\pfrontx,\pfronty)[2]
\seglength=1416  \gaplength=300  
\drawline\scalar[\E\REG](\pbackx,\pbacky)[3]
\drawline\photon[\W\REG](\pfrontx,\pfronty)[5]
\put(34750,9750){$e^-$}
\put(22500,9750){$e^-$}
\put(34650,4650){$H^+$}
\put(34750,1450){$H^-$}
\put(22500,1750){$~\gamma$}
\put(28650,-2000){$(2)$}
\end{picture}
\vskip 3.0cm
\begin{picture}(10000,8000)
\THICKLINES
\bigphotons
\drawline\fermion[\W\REG](24900,10000)[5000]
\drawarrow[\E\ATBASE](\pmidx,\pmidy)
\drawline\fermion[\W\REG](\pbackx,\pbacky)[5000]
\drawarrow[\E\ATBASE](\pmidx,\pmidy)
\drawline\photon[\S\REG](\pfrontx,\pfronty)[9]
\seglength=1416  \gaplength=300  
\drawline\scalar[\NE\REG](\pbackx,\pbacky)[4]
\seglength=1416  \gaplength=300  
\drawline\scalar[\E\REG](\pfrontx,\pfronty)[3]
\drawline\photon[\W\REG](19900,1000)[5]
\put(25400,9750){$e^-$}
\put(13150,9750){$e^-$}
\put(25150,5500){$H^+$}
\put(25400,750){$H^-$}
\put(13150,750){$~\gamma$}
\put(19300,-2000){$(3)$}
\end{picture}
\vskip 3.0cm
\begin{picture}(10000,8000)
\THICKLINES
\bigphotons
\drawline\fermion[\SW\REG](16350,10000)[6000]
\drawarrow[\NE\ATBASE](\pmidx,\pmidy)
\drawline\photon[\SE\REG](\pbackx,\pbacky)[3]
\seglength=1416  \gaplength=300  
\drawline\scalar[\E\REG](\pbackx,\pbacky)[2]
\seglength=1416  \gaplength=300  
\drawline\scalar[\SE\REG](\pfrontx,\pfronty)[2]
\drawline\fermion[\W\REG](\fermionbackx,\fermionbacky)[6000]
\drawarrow[\E\ATBASE](\pmidx,\pmidy)
\drawline\fermion[\NW\REG](\pbackx,\pbacky)[6000]
\drawarrow[\SE\ATBASE](\pmidx,\pmidy)
\drawline\photon[\SW\REG](\pfrontx,\pfronty)[6]
\put(16850,9750){$e^-$}
\put(350,9750){$e^-$}
\put(17700,3500){$H^+$}
\put(16600,1250){$H^-$}
\put(600,1500){$~\gamma$}
\put(8650,-1000){$(4)$}
\THICKLINES
\bigphotons
\drawline\fermion[\SW\REG](38350,10000)[6000]
\drawarrow[\NE\ATBASE](\pmidx,\pmidy)
\drawline\photon[\SW\REG](\pbackx,\pbacky)[16]
\drawline\fermion[\W\REG](\fermionbackx,\fermionbacky)[6000]
\drawarrow[\E\ATBASE](\pmidx,\pmidy)
\drawline\fermion[\NW\REG](\pbackx,\pbacky)[6000]
\drawarrow[\SE\ATBASE](\pmidx,\pmidy)
\drawline\photon[\SE\REG](\pfrontx,\pfronty)[12]
\seglength=1416  \gaplength=300  
\drawline\scalar[\E\REG](\pbackx,\pbacky)[2]
\seglength=1416  \gaplength=300  
\drawline\scalar[\SE\REG](\pfrontx,\pfronty)[2]
\put(38850,9750){$e^-$}
\put(22350,9750){$e^-$}
\put(39250,-2100){$H^+$}
\put(38250,-4650){$H^-$}
\put(22600,-4500){$~\gamma$}
\put(30650,-7000){$(5)$}
\end{picture}
\vskip 3.0cm
\centerline{\bf\Large Fig.4}
\vfill
\end{document}